\def \PR {{ Phys. Rev.} }
\def \PL {{ Phys. Lett.} }

\def \NP {{ Nucl. Phys.} }
\def \PRL {{ Phys. Rev. Lett. }}
\def \bc {\begin{center}}
\def \ec {\end{center}}

\def \bfr {\begin{flushright}}
\def \efr {\end{flushright}}

\def \ba {\begin{array}}
\def \ea {\end{array}}

\def \bea {\begin{eqnarray}}
\def \eea {\end{eqnarray}}

\def \be {\begin{equation}}
\def \ee {\end{equation}}

\def\overpmup {\overleftarrow{\partial}\,\,\!\!\!\!\!\!\!
\overrightarrow{\hbox{\ \ }^\mu}}

\def\ni{\noindent}
\def\nn{\nonumber}
\def\p{\partial}
\def\f{\frac}
\def\l[{\left[}
\def\r]{\right]}
\def\TG{\tilde{G}}
\def\TT{\tilde{T}}

\def\tg{\tilde{g}}

\def\tit{\tilde{g}_t}
\def\tr{\hbox{tr}}
\def\um{\frac{1}{2}}
\def\nlnrs{N^{\lambda\nu\rho\sigma}_{(\pm)}}
\def\nilnrs{N_{\lambda\nu\rho\sigma}^{(\pm)}}
\def\nlnrsp{N^{\lambda\nu\rho\sigma}_{(+)}}

\def\mlnrs{M^{\lambda\nu\rho\sigma}_{(\pm)}}
\def\kpm{\kappa_{(\pm)}}
\def\kmp{\kappa_{(\mp)}}

\newcommand{\coci}[1]{ \frac{i}{r^2}\int{{d}^3x\,{\rm tr}\left[ #1 \right]}}
\newcommand{\xl}[1]{ {\tilde{X}}^{L}_{#1} }

\newcommand{\xr}[1]{ {\tilde{X}}^{R}_{#1} }

\newcommand{\apm}[1]{a^{(\pm)}_{#1}}
\newcommand{\apmb}[1]{\bar{a}^{(\pm)}_{#1}}
\newcommand{\cpm}[1]{c^{(\pm)}_{#1}}
\newcommand{\cpmb}[1]{\bar{c}^{(\pm)}_{#1}}

\newcommand{\cpb}[1]{\bar{c}^{(+)}_{#1}}

\newcommand{\cmb}[1]{\bar{c}^{(-)}_{#1}}
\newcommand{\apb}[1]{\bar{a}^{(+)}_{#1}}
\newcommand{\amb}[1]{\bar{a}^{(-)}_{#1}}
\newcommand{\bpmb}[1]{\bar{b}^{(\pm)}_{#1}}
\newcommand{\bpm}[1]{{b}^{(\pm)}_{#1}}

\newcommand{\fpm}[1]{f^{(\pm)}_{#1}}
\newcommand{\fp}[1]{f^{(+)}_{#1}}
\newcommand{\fm}[1]{f^{(-)}_{#1}}

\newcommand{\fpmb}[1]{\bar{f}^{(\pm)}_{#1}}
\newcommand{\fpb}[1]{\bar{f}^{(+)}_{#1}}
\newcommand{\fmb}[1]{\bar{f}^{(-)}_{#1}}
\newcommand{\chipmb}[1]{\bar{\chi}^{(\pm)}_{#1}}
\newcommand{\chipm}[1]{\chi^{(\pm)}_{#1}}
\newcommand{\hpm}[1]{h^{(\pm)}_{#1}}
\newcommand{\hpmb}[1]{\bar{h}^{(\pm)}_{#1}}

\newcommand{\hmb}[1]{\bar{h}^{(-)}_{#1}}
\newcommand{\A}[1]{A^{(\pm)}_{#1}}
\newcommand{\phipm}[1]{\varphi^{(\pm)}_{#1}}
\documentstyle[12pt,a4]{article}

\begin{document}


\begin{center} 
{\bf GAUGE TRANSFORMATION PROPERTIES OF VECTOR AND
TENSOR POTENTIALS  REVISITED: A GROUP QUANTIZATION
APPROACH}
\footnote{Work partially
supported by the DGICYT.}
\end{center}
\bigskip
\bigskip
\centerline{ {\it M. Calixto$^{1,3}$\footnote{E-mail: 
pymc@swansea.ac.uk / calixto@ugr.es} and  
 V. Aldaya$^{2,3}$\footnote{E-mail: valdaya@iaa.es}} }
\bigskip

\begin{enumerate}
\item {Department of Physics, University of Wales Swansea, Singleton Park, 
Swansea, SA2 8PP, U.K.}
\item {Instituto de Astrof\'{\i}sica de Andaluc\'{\i}a, Apartado Postal 3004,
18080 Granada, Spain.}
\item  {Instituto Carlos I de F\'\i sica Te\'orica y Computacional, Facultad
de Ciencias, Universidad de Granada, Campus de Fuentenueva, 
Granada 18002, Spain.} 
\end{enumerate}

\bigskip
\begin{center}
{\bf Abstract}
\end{center}
\small

\begin{list}{}{\setlength{\leftmargin}{3pc}\setlength{\rightmargin}{3pc}}
\item 

The possibility of 
non-trivial representations of the gauge group on wavefunctionals 
of a gauge invariant quantum field theory leads 
to a generation of mass for intermediate vector and tensor bosons. The mass 
parameters $m$ show up as central charges in the algebra of constraints, which 
then become of second-class nature. The gauge group coordinates 
acquire dynamics outside the null-mass shell and provide the 
longitudinal field degrees of freedom that massless bosons need to form 
massive bosons. This is a {\it non-Higgs} 
mechanism that could provide new clues 
for the best understanding of the symmetry breaking mechanism in unified 
field theories. A unified quantization of massless and massive 
non-Abelian Yang-Mills, 
linear Gravity and Abelian two-form gauge field theories are fully developed 
from this new approach, where a cohomological origin of mass is pointed out.
\end{list}
\ni {\bf Keywords}: algebraic and geometric quantization, group cohomology, 
constraints.
\normalsize

\vskip 1cm

\section{Introduction}

In this paper we discuss a new approach to quantum gauge theories, from 
a group-theoretic perspective, in which mass enters the theory 
in a {\it natural} way. More precisely, the presence of mass will manifest 
through non-trivial responses $U\Psi=D^{(m)}_{\TT}(U)\Psi$  
of the wavefunctional $\Psi$ under the action of gauge transformations 
$U\in\TT$, where we denote by $D^{(m)}_{\TT}$ a specific representation of 
the gauge group $\TT$ with index $m$. The standard case $D^{(m)}_{\TT}(U)=1\,, 
\forall U\in \TT$ corresponds to the well-known `Gauss law' condition, which 
also reads $X_{a}\Psi=0$ for infinitesimal gauge transformations 
$U\sim 1+\varphi^a X_a$. The case of Abelian representations 
$D^{(\vartheta)}_{\TT}(U_n)=e^{in\vartheta}$ of $\TT$, 
where $n$ denotes the winding 
number of $U_n$, leads to the well-known $\vartheta$-vacuum phenomena. 
We shall see that more general (non-Abelian) representations 
$D^{(m)}_{\TT}$ of the gauge group $\TT$ entail 
{\it non-equivalent quantizations} (in the sense of, e.g. 
\cite{McMullan,Landsman}) and a {\it generation of mass}.
 
This non-trivial response of $\Psi$ under gauge transformations $U$ 
causes a {\it deformation} of the corresponding 
Lie-algebra commutators and leads to 
the appearance of central terms proportional to  mass parameters 
(eventually parametrizing the {\it non-equivalent quantizations}) in
the algebra of constraints, which then become a mixture of first- and 
second-class constraints. As a result,  extra (internal) field degrees 
of freedom emerge out of second-class constraints and are 
transferred to the gauge potentials to conform massive  bosons 
(without Higgs fields!).

Thus, the `classical' case $D^{(m)}_{\TT}=1$ is not in general 
preserved in passing to 
the quantum theory. Upon quantization, first-class constraints (connected 
with a gauge invariance of the classical system) might become second-class, a 
metamorphosis which is familiar when quantizing {\it anomalous} gauge 
theories. Quantum ``anomalies'' change the picture of physical states 
being singlets under the constraint algebra. Anomalous (unexpected) 
situations generally go with the standard viewpoint of quantizing 
classical systems; however, these breakdowns, which sometimes are 
inescapable obstacles for canonical quantization,  could be reinterpreted as 
normal (even essential) situations in a wider setting. 
Dealing with constraints directly in the quantum arena, 
this transmutation in the nature of constraints should be naturally allowed, 
as it provides new richness to the quantum theory. 

A formalism 
which intends to place the familiar correspondence (canonical) 
rules of quantum mechanics, $q\rightarrow \hat{q},\,\, p\rightarrow \hat{p}=
-i\hbar\f{\p}{\p q}$, within a rigorous frame is Geometric Quantization (GQ) 
\cite{GQ}. The basic idea in this formalism is that the quantum theory 
should be an irreducible representation of the Poisson algebra of observables 
of the classical phase space. However, 
it is well known that this program cannot 
be fully executed because technical obstructions arise, 
mainly due to ordering 
problems (see, for example, \cite{nogo,symplin}). Some of these limitations 
can be avoided by replacing the phase-space manifold by a group, which 
is the spirit of the Group Approach to Quantization (GAQ) program 
\cite{GAQ}. 
Needless to say, the requirement of an underlying  group structure 
represent some drawback, although less, in practice, than it might 
seem. After all, any consistent (non-perturbative) quantization is nothing  
other than a unitary irreducible representation of 
a suitable (Lie, Poisson) algebra. 
Also, constrained quantization (see below and Refs. \cite{Ramirez,FracHall}) 
increases the range of applicability of the formalism. Nonetheless, we should 
remark that the GAQ formalism, which is at heart an operator 
description of a quantum system, is not meant to quantize a classical system 
(a phase space), but rather a {\it quantizing group} $\TG$ is the primary 
object. Even more, in some cases 
(anomalous groups \cite{chorri,Marmo,virazorro}, 
for example), it is unclear how to associate a definite 
classical phase space with the 
quantum theory obtained, thus weakening
the notion itself of classical limit. 
Furthermore, this {\it cohomological} mechanism 
of mass-generation makes perfect sense from the GAQ 
framework and we are going 
to use its concepts and tools to work out the quantization of 
vector and tensor potentials.

Quantizing on a group  
requires the revision of some standard concepts,  
such as {\it gauge transformations}, in order to deal properly with them. 
The meaning of gauge transformations in Quantum Mechanics is not well 
understood at present (see, for example, \cite{Rovelli}); thus, a 
reexamination of it is timely. 

In a previous article \cite{empro},  a revision of the traditional concept 
of gauge transformation for the electromagnetic vector potential, 
\be 
\varphi(x)\rightarrow \varphi(x)+\varphi'(x),\;\;\;
{\cal A}_{\mu}(x)\rightarrow {\cal A}_{\mu}(x)-\partial_{\mu} \varphi'(x)\,,
\label{connection} 
\ee
\ni was necessary to arrange this transformation inside a group law; that is, 
to adapt this operation to an action of a group on itself: 
the group law of the (infinite-dimensional) 
{\it electromagnetic quantizing group} $\TG$. The proposed Lie group $\TG$
had a principal bundle structure $\TG\rightarrow\TG/\TT$ and was
parameterized, roughly speaking, by the coordinates $A_{\mu}(\vec{x},t)$ of
the Abelian subgroup $G_A$ of Lie-algebra valued vector potentials, 
the coordinates $v=(y_\mu,\Lambda_{\mu\nu})$ (space-time translations 
and Lorentz transformations) of the Poincar\'e group $P$ and the coordinates 
$\varphi(x)$ of the local group
$T\equiv{\rm Map}(\Re^4, U(1))$, which took part of the structure group 
$\TT\sim T\times U(1)$ and 
generalized the standard $U(1)$-phase 
invariance, $\Psi\sim e^{i\alpha}\Psi$, in Quantum Mechanics. 
In this way, the extra
$\TT$-equivariance conditions on wave functions [complex valued functions 
$\Psi(\tg)$ on $\TG$], i.e. 
$\Psi(\tg_t*\tg)\sim\Psi(\tg),\,\,\forall \tg\in \TG, \tg_t\in  \TT$, 
provided the traditional constraints of the theory (we denote by $*$ the 
composition law of $\TG$).

The above-mentioned revision 
was motivated by the fact that the transformation (\ref{connection}) 
is not compatible with a quantizing group $\TG$.  In fact, the general property
$g*e=e*g=g$ for a composition law $g''=g'*g$ of a group $G$ ($e$ denotes
the identity element), precludes the existence of linear terms, 
in the group law $g''^j=g''^j(g'^k,g^l)$ of a 
given  parameter $g^j$ of $G$, 
other than $g'^j$ and $g^j$; that is, near the identity 
we have $g''^j=g'^j+g^j+O(2)$ ---in canonical coordinates. 
Therefore, the group law for 
the field parameter ${\cal A}_{\mu}$ cannot have linear terms in $\varphi$. 
The natural way to address this situation is just to 
choose $A_\mu\equiv {\cal A}_{\mu}-\partial_{\mu} \varphi$, which stays 
unchanged under gauge transformations, and change 
the phase $\zeta=e^{i\alpha}$ 
of the quantum-mechanical wave functional $\Psi(A)$ accordingly, as follows:
\bea
\varphi(x)\rightarrow \varphi(x)+\varphi'(x),\;\;\;
A_\mu(x)\rightarrow A_\mu(x),\nn \\
\zeta\rightarrow \zeta 
\exp\left\{-\frac{i}{2c\hbar^2}\int_\Sigma d\sigma_\mu(x) 
\eta^{\rho\sigma}\partial_\rho \varphi'(x)\overpmup 
A_\sigma(x)\right\}\, ,\label{gaugenew} 
\eea
\ni where  $\eta^{\rho\sigma}$ denotes the Minkowski metric, 
$\Sigma$ denotes a spatial hypersurface and $\hbar$ is the Plank constant,
which is required to kill dimensions of $\partial_\rho\varphi'\overpmup 
A^\rho\equiv\partial_\rho\varphi'
\partial^\mu  A^\rho- A^\rho
\partial^\mu\partial_\rho \varphi'$ and gives a 
{\it quantum} character to the transformation (\ref{gaugenew}) versus 
the {\it classical} character of (\ref{connection}) [from now on, we shall use 
natural unities $\hbar=1=c$]. The piece  $\partial_\rho\varphi'\overpmup 
A^\rho$  in (\ref{gaugenew}) takes part of a {\it symplectic} current 
\be
J^\mu(g'|g)(x)\equiv\um\eta^{\rho\sigma}[(vA')_\rho(x)-
\partial_\rho(v\varphi')(x)]\overpmup [A_\sigma(x)-
\partial_\sigma\varphi(x)]\,, 
\ee
[we are denoting $g\equiv(A,\varphi,v)$ and $(vA')_\rho(x)\equiv 
\f{\p v^\alpha(x)}{\p x^\rho}A'_\alpha(v(x))$, $(v\varphi')(x)\equiv 
\varphi'(v(x))$,  with 
$v^{\alpha}(x)=\Lambda^{\alpha}_\beta x^\beta +y^\alpha$ the general 
action of the restricted Poincar\'e group $P$ on Minkowski space-time] 
which is conserved,
$\partial_\mu J^\mu=0$,  if  $A_\nu$ and $\varphi$ satisfy the field
equations $(\partial_\mu\partial^\mu + m^2) A_\nu=0$ and
$(\partial_\mu\partial^\mu + m^2) \varphi=0$ ($m$ is a parameter with mass
dimension), so that the integral in (\ref{gaugenew}) does not depend on the
chosen spacelike hypersurface $\Sigma$. The integral $\xi(g'|g)\equiv
\int_\Sigma d\sigma_\mu(x)J^\mu(g'|g)(x)$  is a two-cocycle $\xi: G\times
G\rightarrow \Re$ [$G$ denotes the semi-direct product 
$(G_A\times T)\times_v P$],  which fulfills the well-known properties:
\be\ba{cc}
\xi(g_2|g_1)+\xi(g_2*g_1|g_3)=\xi(g_2|g_1*g_3)+\xi(g_1|g_3)\;\;, 
\forall g_i\in G\,,\\ 
\xi(g|e)=0=\xi(e|g)\,,\;\;\;\forall g\in G\,, \ea
\ee
\ni and is the basic ingredient to construct the 
centrally extended group law $\tg''=\tg'*\tg$, more explicitly
\be 
\tg''\equiv (g'';\zeta'')=(g'*g;\zeta'\zeta e^{i\xi(g'|g)})\,,
\;\;\; g,g',g''\in G;\,\,\zeta, \zeta',\zeta''\in U(1)\,,
\ee
of the  electromagnetic quantizing group $\TG$ (see below and Ref.
\cite{empro} for more details). 

It is worth mentioning that
the required review of the concepts of gauge transformations and constraint 
conditions to construct the quantizing group $\TG$ has led, as a byproduct, 
to a unified quantization of both the electromagnetic and Proca fields 
\cite{empro}, within the same general scheme of 
quantization based on a group (GAQ) \cite{GAQ}; clearly, a unified scheme of 
quantization for massless and 
massive gauge theories is suitable as an alternative to the standard 
Spontaneous Symmetry Breaking mechanism, which is intended to supply mass
whereas preserving renormalizability. On the other hand, 
the standard (classical) transformation  (\ref{connection}) is regained 
as the trajectories associated with  generalized equations 
of motion generated by vector fields with null Noether invariants 
({\it gauge subalgebra}, see Refs. \cite{empro,config2} and below). 

This particular revision also applies for the non-Abelian case (Yang-Mills
potentials), and for gravity itself, as they are gauge theories. However, 
for these cases, the situation seems to be a little more subtle and 
complicated. The goal of this article is to present a (non-perturbative) 
group approach to quantization  of 
non-Abelian gauge theories and to point out a cohomological origin of mass, 
as a consequence of a new look to gauge transformations 
and constraints. Furthermore, although a proper group formulation 
of a quantum theory of gravity is beyond the scope of this article, the
corresponding linearized version of this theory can be a useful model 
to work out, as a preliminary step towards the more complicated, 
non-linear case. Even more so, some two-dimensional toy models 
of quantum gravity, such as {\it string inspired} or {\it BF} theories 
\cite{Jackiw123}, also admit a group formulation similar to 
the non-Abelian Yang-Mills theories.

The organization of the present paper is the following. 
In Sec. 2, a quantizing group $\TG$ for linear gravity and 
Abelian two-form gauge theories (symmetric and anti-symmetric tensor 
potentials, respectively) is offered. This simple example contains most 
of the essential elements of more general cases to which 
GAQ is applied and we shall make use of it to explain the method; 
the structure and nature 
of constraints, the count of field degrees of freedom, the 
Hilbert space and the physical operators of these theories 
is presented, for the massless and massive cases, in a unified manner. 
In Sec. 3, a quantizing 
group for general, non-Abelian Yang-Mills gauge theories is proposed; 
the full quantization is then performed inside the GAQ framework, 
and the connection with other approaches to quantization 
is given; also, non-trivial representations of the gauge group $T$ are 
connected with the {\it $\vartheta$-vacuum} phenomenon and with mass 
generation for Yang-Mills vector fields; this is a {\it non-Higgs} 
mechanism which can provide new clues for the best understanding of the 
nature of the symmetry-breaking mechanism. 
Finally, Sec. 4 is devoted to some comments, conclusions 
and outlooks. 

\section{Unified quantization of massless and massive tensor potentials}

Tensor potentials ${\cal A}_{\lambda\nu}(x)$ are primary objects 
for field theories 
such as gravity and two-form gauge theories. More specifically, 
the symmetric and anti-symmetric parts ${\cal A}^{(\pm)}_{\lambda\nu}(x)
\equiv \um [{\cal A}_{\lambda\nu}(x)\pm {\cal A}_{\nu\lambda}(x)]$ of 
${\cal A}_{\lambda\nu}(x)$ are the corresponding 
tensor potentials for linearized gravity and Abelian two-form gauge 
theories, respectively. These theories have a gauge freedom of the form 
(see, for example, Refs. \cite{Weinberg,Teitelboim,Nieuwen}):
\be
\phipm{\lambda}(x)\rightarrow \phipm{\lambda}{}'(x)+\phipm{\lambda}(x),\;\;
{\cal A}^{(\pm)}_{\lambda\nu}(x)\rightarrow 
{\cal A}^{(\pm)}_{\lambda\nu}(x)-\um\left(\partial_\lambda\phipm{\nu}{}'(x)
\pm\partial_\nu\phipm{\lambda}{}'(x)\right)\,,\label{tensor}
\ee
where $\phipm{\lambda}$ are vector-valued functions parametrizing 
the four-dimensional
Abelian local group $T^{(\pm)}={\rm Map}(\Re^4, U(1)\times U(1)\times 
U(1)\times U(1))$. As for the vector potential ${\cal A}_\mu$, 
there is a quantum
version of the transformation (\ref{tensor}) compatible with a 
group law, which is explicitly written as:
\bea
\phipm{\lambda}(x)\rightarrow \phipm{\lambda}(x)+\phipm{\lambda}{}'(x),\;\;\;
\A{\lambda\nu}(x)\rightarrow \A{\lambda\nu}(x),\nn \\
\zeta\rightarrow \zeta 
\exp\left\{-\frac{i}{2}\int_\Sigma d\sigma_\mu(x) \nlnrs 
\partial_\lambda \phipm{\nu}{}'(x)\overpmup \A{\rho\sigma}(x)\right\}\, ,
\label{tensornew} 
\eea 
where we denote $\nlnrs\equiv \eta^{\lambda\rho}\eta^{\nu\sigma}\pm  
\eta^{\lambda\sigma}\eta^{\nu\rho}-\kpm \eta^{\lambda\nu}\eta^{\rho\sigma}$ 
with $\kappa_{(+)}=1$ and $\kappa_{(-)}=0$. The symplectic current for this 
case is (for the moment, let us restrict the theory to the simplified 
situation where $v=e_P=$ the identity of the Poincar\'e subgroup $P$)
\be 
J^\mu_{(\pm)}(g'|g)(x)\equiv\um\nlnrs [\A{\lambda\nu}{}'(x)-
\partial_\lambda\phipm{\nu}{}'(x)]
\overpmup [\A{\rho\sigma}(x)-
\partial_\rho\phipm{\sigma}(x)]\,, 
\ee
which is conserved for solutions of the field equations
$(\partial_\mu\partial^\mu + m^2) \A{\rho\sigma}=0$ and 
$(\partial_\mu\partial^\mu + m^2) \phipm{\rho}=0$. 
The integral of this current 
on an arbitrary spatial hypersurface $\Sigma$ splits up into three 
distinguishable and typical (see later on Sec. 3) cocycles 
\bea
\xi^{(\pm)}_1(g'|g)&=&\um\int_\Sigma d\sigma_\mu \nlnrs
\A{\lambda\nu}{}' \overpmup \A{\rho\sigma}\,,\nn\\ 
\xi^{(\pm)}_2(g'|g)&=&-\frac{1}{2}\int_\Sigma d\sigma_\mu \nlnrs [ 
\partial_\lambda \phipm{\nu}{}'\overpmup \A{\rho\sigma}+ 
\A{\lambda\nu}{}'\overpmup 
\partial_\rho \phipm{\sigma}]\,,\label{3cocys}\\
\xi^{(\pm)}_3(g'|g)&=&\frac{1}{2}\int_\Sigma d\sigma_\mu \nlnrs
\partial_\lambda \phipm{\nu}{}'\overpmup 
\partial_\rho\phipm{\sigma}\nn\\
&=&\frac{m^2}{2}\int_\Sigma d\sigma_\mu [ 
\phipm{\lambda}\overpmup \phipm{}{}^\lambda -\f{\kmp}{m^2} 
\partial_\lambda\phipm{}{}^\lambda
\overpmup \partial_\rho\phipm{}{}^\rho]\,.\nn
\eea
The first cocycle  $\xi_1$ is meant to provide {\it dynamics} 
to the tensor potential, so that the couple $(\A{},\dot{A}^{(\pm)})$ 
corresponds to a canonically-conjugate pair of variables; 
the second cocycle $\xi_2$, 
the {\it mixed} cocycle, provides a non-trivial (non-diagonal) action 
of constraints on tensor potentials and determines 
the number of degrees of freedom of the constrained theory [it is also 
responsible for the transformation appearing in the second line of 
(\ref{tensornew})]; the third cocycle $\xi_3$, 
the {\it mass} cocycle,   determines the structure of constraints (first- or 
second-class) and modifies the dynamical content of the
tensor potential coordinates $\A{}$ in the massive case $m\not=0$ by  
transferring degrees of freedom between the 
$\A{}$ and $\phipm{}$ coordinates, thus conforming the massive field. 
In this way, the appearance of mass in the theory 
has a {\it cohomological origin}.  To make more explicit the intrinsic 
significance of these three quantities $\xi_j\,,\,\, j=1,2,3$, let us 
construct the quantum theory of these (anti-)symetric tensor potentials for 
the massless and massive cases in a unified manner, the physical 
interpretation of which will be a free theory of massless and massive 
spin 2 particles (gravitons) for the symmetric case, and a free 
theory of massless spin 0 
pseudo-scalar particles, and massive spin 1 pseudo-vector particles for 
the anti-symmetric case. 

The starting point 
will be the {\it Tensor quantizing group} $\TG^{(\pm)}=\{
\tg=(\A{},\phipm{},v;\zeta)\}$  with group law $\tg''=\tg'*\tg$, which 
can be explicitly written as:
\bea
v''&=&v'*v\,,\;\;\;\;v,v',v''\in P\,,\nn\\
\A{\lambda\nu}{}''(x)&=& (v\A{}{}')_{\lambda\nu}(x)+\A{\lambda\nu}(x)\,,\nn\\
\phipm{\rho}{}''(x)&=& (v\phipm{}{}')_{\rho}(x)+\phipm{\rho}(x)\,,
\label{ley1}\\
\zeta''&=&\zeta'\zeta\exp\left\{i\sum_{j=1}^3\xi^{(\pm)}_j(\A{}{}',
\phipm{}{}',v'|\A{},\phipm{},v)\right\}\,,\nn
\eea
where we denote $(v\A{})_{\lambda\nu}(x)\equiv 
\f{\p v^\alpha(x)}{\p x^\lambda}\f{\p v^\beta(x)}{\p x^\nu}
\A{\alpha\beta}(v(x))$ and so on. 
The entire group $\TG$ will be regarded as a principal fibre bundle 
$\TG\rightarrow\TG/\TT$, with structure 
group $\TT\rightarrow\TT/U(1)$ 
[in general, a non-trivial, central extension $\TT$ of $T$ by $U(1)$], and 
it will be the driver of the quantization procedure. 

The Hilbert space ${\cal H}(\TG)$ 
of the theory will be made of complex $\TT$-equivariant 
functions $\Psi(\tg)$ on $\TG$ (wave functionals), i.e. 
\bea
\Psi: \TG\rightarrow C\,, \;\;\;{\rm such \ that} \;\;\;
\Psi(\tit *\tg)=
D_{\TT}^{(\epsilon)}(\tit)\Psi(\tg)\,,\;\;\forall 
\tit\in \TT,\,\,\forall
\tg\in \TG\,,\label{tequiv}
\eea
\ni where $D_{\TT}^{(\epsilon)}$ symbolizes a specific 
representation $D$ of $\TT$ with 
$\epsilon$-index (the mass $\epsilon=m$ or, in particular, 
the {\it $\vartheta$-angle} 
\cite{Jackiwtheta} 
of non-Abelian gauge theories; see below). On the other hand, the 
operators will be the, let us say, 
right-invariant vector fields $\xr{\tg^j}$; 
that is, the generators of the finite left-action of $\TG$ on itself 
$L_{\tg'}(\tg)=\tg'*\tg$ [the corresponding finite right-action
$R_{\tg}(\tg')=\tg'*\tg$  is generated by the left-invariant vector fields
$\xl{\tg^j}$].  

Let us use a Fourier-like parametrization of $\TG$ according to  
the standard decomposition of the field into negative, 
and positive, frequency parts:
\bea
\A{\lambda\nu}(x)\equiv \int d\Omega_k[\apm{\lambda\nu}(k)e^{-ikx}+
\apmb{\lambda\nu}(k)e^{ikx}]\,,\nn\\
\phipm{\rho}(x)\equiv \int d\Omega_k[ \fpm{\rho}(k)e^{-ikx}+
 \fpmb{\rho}(k)e^{ikx}]\,,\label{fourierpar}
\eea
where $d\Omega_k\equiv \f{d^3k}{2k^0}$ is the standard integration measure 
on the positive sheet of the mass hyperboloid $k^2=m^2$. 
The non-trivial Lie-algebra commutators of the left-invariant 
vector fields $\xl{\tg^j}$ are easily  computed from  the group 
law (\ref{ley1}) in the parametrization (\ref{fourierpar}), 
and they are explicitly (for simplicity, we discard the Lorentz subgroup):
\bea
\left[\xl{\apmb{\lambda\nu}(k)},
\xl{\apm{\rho\sigma}(k')}\right]&=& 
 i\nlnrs\Delta_{kk'}\Xi\,, \nn \\ 
\left[\xl{\apm{\lambda\nu}(k)},\xl{\fpmb{\sigma}(k')}\right]&=&
k_{\rho}\nlnrs\Delta_{kk'}\Xi\,,\;\;\;\;
 \left[\xl{\apmb{\lambda\nu}(k)},
\xl{\fpm{\sigma}(k')}\right]=k_{\rho}\nlnrs\Delta_{kk'}\Xi\; ,\nn\\ 
 \left[\xl{\fpmb{\rho}(k)},
\xl{\fpm{\sigma}(k')}\right]&=&ik^2(\eta^{\rho\sigma}-
\kmp\f{k^\rho k^\sigma}{k^2})\Delta_{kk'}\Xi\,, \label{conmutadores1}\\ 
\left[\xl{y^{\mu}},\xl{\apm{\lambda\nu}(k)}\right]&=&
ik_{\mu}\xl{\apm{\lambda\nu}(k)}\,,\;\;\;\; 
\left[\xl{y^{\mu}},\xl{\apmb{\lambda\nu}(k)}\right] =
- ik_{\mu}\xl{\apmb{\lambda\nu}(k)}\,,\nn\\ 
\left[\xl{y^{\mu}},\xl{ \fpm{\rho}(k)}\right]&=&
ik_{\mu}\xl{\fpm{\rho}(k)}\,,\;\;\;\;
\left[\xl{y^{\mu}},\xl{\fpmb{\rho}(k)}
\right]=-ik_{\mu}\xl{\fpmb{\rho}(k)}\,,\nn
\eea
where $\Delta_{kk'}=2k^0\delta^3(k-k')$ is the generalized delta function 
on the positive sheet of the mass hyperboloid, and we denote by 
$\Xi\equiv i\xl{\zeta}=i\xr{\zeta}$ the central generator, 
in order to distinguish it from 
the rest, in view of its  ``central'' (important) 
role in the quantization procedure; it   
behaves as $i$ times the identity operator, $\Xi\Psi(\tg)=i\Psi(\tg)$, 
when the $U(1)$ part of the 
$\TT$-equivariance conditions (\ref{tequiv}), 
$D_{\TT}^{(\epsilon)}(\zeta)=\zeta$ 
(always faithful, except in the classical limit $U(1)\rightarrow \Re$ 
\cite{GAQ}), is imposed. 

The representation $L_{\tg'}\Psi(\tg)=\Psi(\tg'*\tg)$ of $\TG$ on wave 
functions $\Psi$ proves to be
reducible. The reduction can be achieved by means of those right-conditions 
$R_{\tg}\Psi(\tg')=\Psi(\tg'*\tg)\equiv\Psi(\tg')$ 
(which commute with the left-action $L$) 
compatible with the above $U(1)$-equivariant  condition $\Xi\Psi=i\Psi$,  
i.e., by means of the so-called {\it polarization conditions} 
$\xl{ }\Psi=0$ \cite{GAQ}. Roughly speaking, a polarization  corresponds 
to a maximal left-subalgebra ${\cal G}_p$ of 
the Lie algebra $\tilde{\cal G}^L$ of $\TG$ which exclude 
the central generator $\Xi$; or, in other words, a maximal, 
{\it horizontal}   
left-subalgebra ${\cal G}_p$. The horizontality property for a subalgebra 
${\cal G}_H=<\xl{\tg^j}>$  can be formally stated as 
$\Theta(\xl{\tg^j})=0,\,\,\forall \xl{\tg^j}\in {\cal G}_H$, 
where we denote by $\Theta$ the 
``vertical'' component $\tilde{\theta}^{L(\zeta)}$ 
[dual to $\Xi=i\xl{\zeta}$, i.e., $\tilde{\theta}^{L(\zeta)}(\xl{\tg^j})=
\delta_{\tg^j}^\zeta$] 
of the standard canonical (Lie-valued) 
left-invariant 1-form $\tilde{\theta}^L$ of $\TG$. It can be easily 
calculated from the general expression:
\be 
\Theta=\left.\f{\p}{\p g^j}\xi(g'|g)\right|_{g'=g^{-1}}dg^j
-i\zeta^{-1}d\zeta\,.\label{thetagen}
\ee
For cases such as the centrally-extended Galilei group \cite{GAQ}, 
the so called {\it quantization 1-form}  $\Theta$ reduces to the 
Poincar\'e-Cartan form of Classical Mechanics, 
$\Theta_{PC}=pdq-\f{p^2}{2m}dt$, 
except for the (typically quantum) phase term  $-i\zeta^{-1}d\zeta$. In the 
same way as $\Theta_{PC}$, the quantization 1-form $\Theta$ gives the 
generalized {\it classical} equations of motion of the system. The 
trajectories for (\ref{thetagen}) are given by the integral curves of the 
characteristic module ${\rm Ker}\Theta\cap{\rm Ker}d\Theta=
\{{\rm vector \ fields} \,\tilde{X}\,/ \, \Theta(\tilde{X})=0,\,
d\Theta(\tilde{X})=0\}$ of the 
{\it presymplectic} (in general, it has a non-trivial kernel) two-form 
$d\Theta$ on the group $\TG$.
The characteristic module of $\Theta$ is generated by the  
{\it characteristic subalgebra} ${\cal G}_c$, which includes 
non-symplectic (non-dynamical) generators; that is, horizontal 
left-invariant vector 
fields which, under commutation, do not give rise to central terms 
proportional to $\Xi$ (i.e., they do not have any  conjugated 
counterpart). In fact, $d\Theta$ at the identity of the group regains the
Lie-algebra cocycle $\Sigma$:
\be
\Sigma\equiv d\Theta|_e:{\cal G}^L\times {\cal G}^L\rightarrow\Re\,,\;\;
\Sigma(X_{a}^L,X_{b}^L)=d\Theta(X_{a}^L,X_{b}^L)|_e=
\Theta(\left[\tilde{X}_{a}^L,\tilde{X}_{b}^L\right])|_e\label{Sigma}
\ee

A glance at the commutators (\ref{conmutadores1}) tells us 
the content of the characteristic subalgebra for $\TG$:
\be
{\cal G}^{(\pm)}_c=<\xl{y^\mu},\,(\xl{\Lambda^\mu_\nu}),\,
\xl{\hpm{\rho}(k)},\,\xl{\hpmb{\sigma}(k)}>
\,\,\forall k\,;
\ee
That is, ${\cal G}_c$ contains the whole Poincar\'e subalgebra 
(when the Lorentz transformations $\Lambda$ are kept) 
and two particular combinations
\be
{\cal G}^{(\pm)}_{{\small{\rm gauge}}}= <\xl{\hpm{\rho}(k)}\equiv 
\xl{\fpm{\rho}(k)}+ik_\lambda
\xl{\apm{\lambda\rho}(k)}\,,\,\,
\xl{\hpmb{\sigma}(k)}\equiv \xl{\fpmb{\sigma}(k)}-ik_\lambda
\xl{\apmb{\lambda\sigma}(k)}>\,,\label{gaugesub}
\ee
which define the {\it gauge subalgebra} ${\cal G}_{{\small{\rm gauge}}}$ 
of the theory. Let us justify this name for 
${\cal G}_{{\small{\rm gauge}}}$. As already commented, the 
trajectories associated with 
vector fields  $\xl{}\in{\cal G}_c$ represent the generalized 
{\it classical} equations of motion (they generalize the standard classical 
equations of motion because they contain also the evolution of the phase 
parameter $\zeta$); for example, the flow of 
$\xl{y^\mu}$ represents the 
space-time evolution $\apm{\lambda\nu}(k)\rightarrow 
e^{-iky}\apm{\lambda\nu}(k)$, whereas the flow of (\ref{gaugesub}),  
\be
\fpm{\lambda}(k)\rightarrow\fpm{\lambda}(k)+\hpm{\lambda}(k)\,,\;\;\;  
\apm{\lambda\nu}(k)\rightarrow\apm{\lambda\nu}(k)+\f{i}{2}
[k_\lambda\hpm{\nu}(k)\pm k_\nu\hpm{\lambda}(k)]\,,
\ee
(and the complex-conjugated counterpart) recovers the classical 
gauge transformations (\ref{tensor}) in Fourier coordinates. 
The invariant quantities 
under the above-mentioned classical equations of motion are 
$F_{\tg^j}\equiv i_{\xr{\tg^j}}\Theta=\Theta(\xr{\tg^j})$, which define 
the (generalized) Noether invariants of the theory 
(total energy, momentum, initial conditions, etc.). 
It can be seen that 
the Noether invariants associated with gauge vector-fields are identically 
zero $F_{\hpm{\lambda}(k)}=0=F_{\hpmb{\lambda}(k)}$. 
Even more, in general, 
the gauge subalgebra ${\cal G}_{{\small{\rm gauge}}}$ constitutes 
a horizontal {\it ideal} of the whole Lie algebra $\tilde{{\cal G}}$ of $\TG$ 
(a {\it normal} horizontal subgroup $N$ for finite transformations) for 
which the right-invariant vector fields can be 
written as a linear combination 
of the corresponding left-invariant vector fields. In fact, the coefficients 
of this linear combination provide a representation of the complement 
of ${\cal G}_{{\small{\rm gauge}}}$ in ${\cal G}_c$ (in this case, 
the Poincar\'e subalgebra). All these properties characterize 
a gauge subgroup of $\TG$. 

Note the subtle distinction between {\it gauge symmetries} and {\it 
constraints} inside the GAQ framework. Constraints take part of the 
structure group $\TT$ and have a non-trivial action 
(\ref{tequiv}) on wave functions. The Lie algebra of $\TT$,
\be
\tilde{{\cal T}}^{(\pm)}=<\xl{\fpm{\rho}(k)}\,,\,\,
\xl{\fpmb{\sigma}(k)}\,,\,\,\Xi>\,,\label{constraints}
\ee
is not a horizontal ideal but, rather, $\TT$ is itself a 
quantizing group [more precisely, a central extension of $T$ by $U(1)$] 
with its own quantization 1-form $\Theta_{\TT}$. {\it First-class  
constraints} will be defined as the characteristic subalgebra   
${\cal T}_c$ of $\tilde{\cal T}$ with respect to $\Theta_{\TT}$. 
{\it Second-class constaints} are defined as the 
complement of ${\cal T}_c$ in $\tilde{\cal T}$, and can be arranged 
into couples of conjugated generators. The constrained Hilbert space 
${\cal H}_{{\small{\rm phys.}}}$ will be made of complex 
wave functionals (\ref{tequiv}) which are annihilated  
\be
\xr{\tg_t}\Psi_{{\small{\rm phys.}}}=0,\,\,\,\,\forall \xr{\tg_t}\in 
{\cal T}_p\label{hwc}
\ee
by a polarization subalgebra 
${\cal T}_p\subset \tilde{\cal T}$ of right-invariant vector fields, 
which contains ${\cal T}_c$ and `half' of second-class constraints 
(the, let us say, `positive modes'). The algebra ${\cal T}_p$ is then the 
maximal right-subalgebra of $\tilde{\cal T}$ that can be consistently 
imposed to be zero on wavefunctionals as constraint equations. The condition 
(\ref{hwc}) selects those wave functionals 
$\Psi_{{\small{\rm phys.}}}$ which transform as  
`highest weight vectors' under $\TT$.
 
Since constraint conditions (\ref{hwc}) are imposed as infinitesimal 
right-restictions (finite left-restrictions), it 
is obvious that not all the operators $\xr{}$ will preserve the constraints;  
we shall call $\tilde{{\cal G}}_{\small{\rm good}}\subset\tilde{{\cal G}}^R$ 
the subalgebra  of ({\it good}$\sim$physical) operators which do so. These 
have to be found inside the {\it normalizer} of the constraints; for example, 
 a sufficient condition for $\tilde{{\cal G}}_{\small{\rm good}}$ 
to preserve the constraints is:
\be
\left[\tilde{{\cal G}}_{\small{\rm good}},{\cal T}_p\right]\subset {\cal T}_p
\,.\label{gooddef}
\ee
From this characterization, we see that first-class constraints 
${\cal T}_c$ become a horizontal ideal (a gauge subalgebra) 
of $\tilde{{\cal G}}_{\small{\rm good}}$, which now defines the constrained 
theory. Even more, it can be proved that gauge generators 
(\ref{gaugesub}) are trivial (zero) on polarized wave functions (see later).

With all this information at hand, let us go back to the 
reduction process of the representation (\ref{tequiv}). An operative 
method to obtain a {\it full} polarization subalgebra ${\cal G}_p$  
is to complete the characteristic subalgebra ${\cal G}_c$ to a 
maximal, horizontal left-subalgebra. Roughly speaking, ${\cal G}_p$ will 
include non-symplectic generators in ${\cal G}_c$ and 
half of the symplectic ones (either ``positions'' or ``momenta''). 
The above-mentioned polarization conditions 
\be
\xl{\tg_p}\Psi=0,\,\,\forall \xl{\tg_p}\in {\cal G}_p \label{polcond}
\ee
[or its finite counterparts $R_{\tg_p}\Psi(\tg)=\Psi(\tg)$]  
represent the generalized {\it quantum} equations of motion; 
for example, $\xl{y^0}\Psi=0$ represents the Schr\"odinger equation. 
In this way, the concept of polarization here generalizes the analogous 
one in Geometric Quantization \cite{GQ}, where no characteristic module 
exists (since all variables are symplectic). 
The content of ${\cal G}_p$ will depend on the value of $k^2$, as also 
happens for ${\cal T}_p$. From now on 
we shall distinguish between the cases $k^2=0$ and $k^2=m^2\not=0$, and 
between symmetric and antis-ymmetric tensor potentials, placing each one in 
separate subsections. Let us see how to obtain the corresponding constrained 
Hilbert space and the action of the physical operators on wave functions.

\subsection{$\TG(k^2=0)$: Massless tensor fields}

Firstly, we shall consider the massless case. A polarization subalgebra for 
the symmetric and anti-symmetric cases is:
\be
{\cal G}_p^{(\pm)}=< {\cal G}_c\,,\,\,\,\xl{\apm{\lambda\nu}(k)}>
\,\,\forall k\,.
\ee
The corresponding $U(1)$-equivariant, polarized wave 
functions (\ref{polcond}) have the following general form:
\bea
\Psi^{(\pm)}\left(y,(\Lambda),\apm{},\apmb{},\fpm{},\fpmb{},\zeta\right)&=& 
W^{(\pm)}\cdot\Phi^{(\pm)}\left(\cpmb{\lambda\nu}(k)e^{-iky}\right)\,,\nn\\
W^{(\pm)}&=& \zeta\exp\left\{\um\int{ d\Omega_k\nlnrs 
\cpmb{\lambda\nu}(k)
\cpm{\rho\sigma}(k)}\right\}\,,\label{wavetensor}\\
\cpmb{\lambda\nu}&\equiv&\apmb{\lambda\nu}+\f{i}{2}(k_\lambda\fpmb{\nu}
\pm k_\nu\fpmb{\lambda})\nn
\eea
where $\Phi$ is an arbitrary power series on its arguments. 
For example, the zero-order wave function (the {\it vacuum}), 
and the one-particle states of momentum $k$ are $|0\rangle\equiv 
W$ and $\hat{a}^{(\pm)\dag}_{\lambda\nu}(k)
|0\rangle \equiv \f{1}{4}\nilnrs\xr{\apm{\rho\sigma}(k)}|0\rangle= 
W^{(\pm)}\cdot[\cpmb{\lambda\nu}(k)e^{-iky}]$, respectively. 
The last equality defines the 
creation operators of the theory whereas the corresponding annihilation 
operators are $\hat{a}^{(\pm)}_{\lambda\nu}(k)\equiv 
\f{1}{4}\nilnrs\xr{\apmb{\rho\sigma}(k)}$ and its 
action on polarized wave functions (\ref{wavetensor}) is 
$\hat{a}^{(\pm)}_{\lambda\nu}(k)\Psi^{(\pm)}=W^{(\pm)}
\f{\delta\Phi^{(\pm)}}{\delta\cpmb{\lambda\nu}(k)}$. One can easily check 
that the action of the gauge operators [the right version of 
(\ref{gaugesub})] $\xr{\hpm{\rho}(k)}$ and 
$\xr{\hpmb{\rho}(k)}$ on polarized wave functions (\ref{wavetensor}) is 
trivial (zero), since they close a {\it horizontal ideal} 
of $\tilde{{\cal G}}$.

The representation 
$L_{\tg'}\Psi(\tg)=\Psi(\tg'*\tg)$ of $\TG$ 
on polarized wave functions (\ref{wavetensor}) is irreducible 
and unitary with respect to the natural scalar product, 
\bea 
\langle \Psi|\Psi'\rangle &=&\int_{\TG}{\mu(\tg)\bar{\Psi}(\tg)\Psi'(\tg)}
\,,\\
 \mu^{(\pm)}(\tg)&=&\tilde{\theta}^{L(1)}\wedge
\stackrel{{\rm dim}(\TG^{(\pm)})}{\dots}
\wedge\tilde{\theta}^{L(n)}=\mu_P\wedge\nlnrs\prod_k d\,
{\rm Re}[\cpm{\lambda\nu}(k)]
\wedge\,d\,{\rm Im}[\cpm{\rho\sigma}(k)]\,,\nn
\eea
where $\mu_P$ means the standard left-invariant 
measure of the Poincar\'e subgroup 
[exterior product $\wedge$ of the components $\tilde{\theta}^{L(j)}$ of 
the standard (Lie-valued) left-invariant 1-form $\tilde{\theta}^L$ of the 
corresponding group]. The finiteness of this scalar product 
is ensured when restricting to constrained wave functions (\ref{tequiv}). 
Before imposing the rest of 
$\TT$-equivariant conditions [the $U(1)$ part has already been imposed], 
we have to look carefully at the 
particular fibration of the structure 
group $\TT\rightarrow \TT/U(1)$ by $U(1)$ in order to separate  
first- from  second-class constraints. A look at the right-version 
of the third line in 
(\ref{conmutadores1}) tells us that all constraints are first-class for the 
massless, symmetric case, whereas the massless, anti-symmetric case 
possesses  a couple of second-class constraints:
\be
\left[\check{k}_\rho\xr{\fmb{\rho}(k)},
\check{k}'_\sigma\xr{\fm{\sigma}(k')}\right]=4i(k^0)^4\Delta_{kk'}\Xi\,,
\label{2ndclass}
\ee
where $\check{k}_\rho\equiv k^\rho$. Thus, first-class constraints for the 
massless anti-symmetric case are ${\cal T}^{(-)}_c=
<\epsilon^\mu_\rho(k)\xr{\fmb{\rho}(k)},\,
\epsilon^\mu_\rho(k)\xr{\fm{\rho}(k)}>,\,
\mu=0,1,2,\,$ where $\epsilon^\mu_\rho(k)$ is a tetrad which diagonalizes the 
matrix $P^{\rho\sigma}=k^\rho k^\sigma$; in particular, we choose 
$\epsilon^3_\rho(k)\equiv \check{k}_\rho$ and 
$\epsilon^0_\rho(k)\equiv k_\rho$.

The constraint equations for the massless, symmetric case are:
\be\ba{rcl}
\xr{\fpb{\sigma}(k)}\Psi^{(+)}_{{\small{\rm phys}}}=0 
&\Rightarrow& \left(2k_\lambda
\f{\delta}{\delta\cpb{\lambda\sigma}(k)}-k^\sigma\eta_{\lambda\nu}
\f{\delta}{\delta\cpb{\lambda\nu}(k)}\right)
\Phi^{(+)}_{{\small{\rm phys}}}=0\,,\\ 
\xr{\fp{\sigma}(k)}\Psi^{(+)}_{{\small{\rm phys}}}=0 
&\Rightarrow& \left(2k^\lambda
\apb{\lambda\sigma}(k)-k^\sigma\eta^{\lambda\nu}
\apb{\lambda\nu}(k)\right)\Psi^{(+)}_{{\small{\rm phys}}}=0\,.
\ea\label{tsym}\ee
The first condition in (\ref{tsym}) says that, in particular, an arbitrary 
combination of one-particle states of momentum $k$, 
$\varepsilon^{\lambda\nu}(k)
\hat{a}^{(+)\dag}_{\lambda\nu}(k)|0\rangle$, is {\it physical} (observable) 
if $2k^\lambda\varepsilon_\lambda^\sigma(k)=k^\sigma
\varepsilon^\lambda_\lambda(k)$. This condition also guarantees that 
physical states have positive (or null) norm, since 
$-\bar{\varepsilon}_{\lambda\nu}\nlnrsp\varepsilon_{\rho\sigma}\geq 0$. 
The second condition in (\ref{tsym}) eliminates {\it null norm vectors} 
from the theory, since it establishes that the physical wave functions  
$\Psi^{(+)}_{{\small{\rm phys}}}$ have support only on the surface 
$2k^\lambda\apb{\lambda\sigma}(k)-k^\sigma\eta^{\lambda\nu}
\apb{\lambda\nu}(k)=0$. In summary, the 8 independent 
conditions (\ref{tsym}) 
keep {\it two} field degrees of freedom out of the original 
10 field degrees of freedom of the symmetric tensor potential. 
The good (physical) operators (\ref{gooddef}) of the theory  are:
\be
\tilde{{\cal G}}^{(+)}_{\small{\rm good}}=<\varepsilon^{\lambda\nu}(k)
\hat{a}^{(+)\dag}_{\lambda\nu}(k),\,\bar{\varepsilon}^{\lambda\nu}(k)
\hat{a}^{(+)}_{\lambda\nu}(k),\,\xr{y^\mu},\,\xr{\Lambda^{\mu\nu}},\,\Xi>\,,
\ee
where the factors $\varepsilon^{\lambda\nu}(k)$ are restricted by the 
above-mentioned conditions [note that ${\cal T}^{(+)}=
\tilde{{\cal T}}^{(+)}/U(1)$ becomes a horizontal 
ideal (gauge subalgebra) of $\tilde{{\cal G}}^{(+)}_{\small{\rm good}}$]. 
The transformation properties of physical 
one-particle states under the Poincar\'e group $P$ declare, in particular, 
that the symmetric tensor field carries helicity $\pm 2$, as corresponds 
to a massless spin 2 particle (graviton).

For the massless anti-symmetric case, only a polarization subalgebra 
${\cal T}_p^{(-)}$ of 
$\tilde{{\cal T}}^{(-)}$ can be consistently imposed as constraint 
equations, due to the non-trivial fibration of $\TT^{(-)}$ by $U(1)$. These 
constraint equations are:
\be\ba{rcl}
\epsilon^\mu_\sigma(k)\xr{\fmb{\sigma}(k)}\Psi^{(-)}_{{\small{\rm phys}}}=0 
&\Rightarrow& k_\lambda\epsilon^\mu_\sigma(k)
\f{\delta}{\delta\cmb{\lambda\sigma}(k)}
\Phi^{(-)}_{{\small{\rm phys}}}=0\,,\;\;\;\mu=0,1,2,3\\ 
\epsilon^\mu_\sigma(k)\xr{\fm{\sigma}(k)}\Psi^{(-)}_{{\small{\rm phys}}}=0 
&\Rightarrow& k^\lambda\epsilon^{\mu\nu}(k)\amb{\lambda\nu}(k)
\Psi^{(-)}_{{\small{\rm phys}}}=0\,,\;\;\;\mu=0,1,2.
\ea\label{tasym}\ee
Only 3 of the four conditions in the first line of (\ref{tasym}) are 
independent, since the combination $\epsilon^0_\sigma\xr{\fmb{\sigma}}$ 
coincides with the gauge operator $k_\sigma\xr{\hmb{\sigma}}$, which 
is identically zero on polarized wave functions (\ref{wavetensor}). 
The second-class character of the constraints (\ref{2ndclass}) precludes 
the imposition of the combination $\mu=3$ in the second line of 
(\ref{tasym}),  so that we have only 2 additional independent conditions 
which, together with the 3 previous ones, keep {\it one} field degree 
of freedom out of the 6 original ones corresponding to 
an anti-symmetric tensor potential. As for the symmetric case, the behaviour   
of the physical states $\Psi^{(-)}_{{\small{\rm phys}}}$ 
under the action of the Poincar\'e group declares that the constrained theory 
corresponds to a pseudo-scalar particle. 
The computation of the good operators 
follows the same steps as for the symmetric case.

\subsection{$\TG(k^2\not=0)$: Massive tensor fields}

As mentioned above, a remarkable characteristic of the quantizing group 
law (\ref{ley1}) is that it accomplishes the quantization of both the 
massless and massive cases, according to the value of the central extension 
parameter $m$ in the third cocycle of (\ref{3cocys}), in a unified way. The 
modification of $\xi^{(\pm)}_3$ in the $m\not=0$ case causes a transfer of 
degrees of freedom between the $\A{}$ and $\phipm{}$ coordinates, so that 
it is possible to decouple the tensor potential by means of a transformation 
which diagonalizes the cocycle $\xi^{(\pm)}$. In fact, the combinations:
\be\ba{rcl}
\xl{\bpmb{\lambda\nu}(k)}&\equiv& \xl{\apmb{\lambda\nu}(k)} +\f{i}{k^2} 
\left(k^\lambda\xr{\fpmb{\nu}(k)}\pm k^\nu\xr{\fpmb{\lambda}(k)} 
-i\eta^{\lambda\nu}k_\alpha k_\beta \xr{\apmb{\alpha\beta}(k)}\right)\,,\\
 \xl{\bpm{\lambda\nu}(k)}&\equiv& \xl{\apm{\lambda\nu}(k)} -\f{i}{k^2} 
\left(k^\lambda\xr{\fpm{\nu}(k)}\pm k^\nu\xr{\fpm{\lambda}(k)} 
+i\eta^{\lambda\nu}k_\alpha k_\beta \xr{\apm{\alpha\beta}(k)}\right)\,,\ea
\ee
commute with the constraints (\ref{constraints}) and close the Lie subalgebra 
\be
\left[\xl{\bpmb{\lambda\nu}(k)},
\xl{\bpm{\rho\sigma}(k')}\right]=
 i\mlnrs(k)\Delta_{kk'}\Xi\,, 
\ee
where $\mlnrs\equiv M^{\lambda\rho}M^{\nu\sigma}\pm  
M^{\lambda\sigma}M^{\nu\rho}-2\kpm M^{\lambda\nu}M^{\rho\sigma}$ 
and $M^{\lambda\rho}(k)\equiv \eta^{\lambda\rho}-
\f{k^\lambda k^\rho}{k^2}$. The different cohomological structure of 
the quantizing group $\TG$ for the present massive case, with regard 
to the abovementioned massless case, leads to a different 
polarization subalgebra and a new structure for constraints. 
The polarization subalgebra is made of the 
following left generators:
\be
{\cal G}_p^{(\pm)}=< {\cal G}_c\,,\,\,\xl{\bpm{\lambda\nu}(k)}\,,\,
\xl{\fpmb{\rho}(k)}>
\,\,\forall k\,.
\ee
The integration of the polarization conditions (\ref{polcond}) on 
$U(1)$-equivariant wave functions leads to 
\bea
\Psi^{(\pm)}&=&\zeta\exp\left\{\um\int{ d\Omega_k\left(\mlnrs 
\bpmb{\lambda\nu}\bpm{\rho\sigma}+k^2M_{(\pm)}^{\lambda\rho}
\chipmb{\lambda}\chipm{\rho}\right)}\right\}\nn\\
& &\cdot\Phi^{(\pm)}\left([\bar{\omega}^{\alpha\beta}_{ij}(k)^{(\pm)}
\bpmb{\alpha\beta}(k)]e^{-iky},\,[\varpi^{\sigma}_l(k)^{(\pm)}
\chipm{\sigma}(k)]e^{iky}\right)\,,\label{wavemass}\\ 
\apm{\lambda\nu}(k)&\equiv&\bpm{\lambda\nu}(k)+\eta^{\alpha\beta}\f{k_\lambda 
k_\nu}{k^2}\bpm{\alpha\beta}(k),\,\,\, \fpm{\lambda}(k)\equiv 
\chipm{\lambda}(k)-2i\f{k_\nu}{k^2}\bpm{\lambda\nu}(k)\,,\nn
\eea
where $M_{(\pm)}^{\lambda\rho}(k)\equiv \eta^{\lambda\rho}-
\kmp\f{k^\lambda k^\rho}{k^2}$;  
$\omega^{\alpha\beta}_{\lambda\nu}(k)^{(\pm)}$ and 
$\varpi^{\sigma}_\lambda(k)^{(\pm)}$ are matrices which diagonalize 
$\mlnrs(k)$ and $M_{(\pm)}^{\lambda\rho}(k)$, respectively, and the indices 
$i,j=1, 2, 3$ and $l=0(+), 1, 2, 3$ label the corresponding eigenvectors 
with non-zero eigenvalue ($l=0(+)$ means ``only for the symmetric case''). 
From the third line of (\ref{wavemass}), we realize the already mentioned 
transfer of degrees of freedom between the $\apm{}$ and $\fpm{}$ coordinates 
for the massive case, leading to a new set of variables, 
$\bpm{}$ and $\chipm{}$, which 
correspond to a massive (anti-)symmetric tensor field and some sort of 
vector field with {\it negative} energy.

For the present massive case, all constraints are second-class for the 
symmetric case, since they close an electromagnetic-like  subalgebra [see 
third line of Eq. (\ref{conmutadores1})], whereas, for the 
anti-symmetric case,  constraints close a Proca-like subalgebra which 
leads to three couples of second-class 
constraints, $\{\bar{\varpi}^{j(-)}_\lambda\xr{\fmb{\lambda}},\,
\varpi^{j(-)}_\lambda\xr{\fm{\lambda}}\}\,\,\,j=1,2,3$, and a couple 
of gauge vector fields $\{k_\lambda\xr{\fmb{\lambda}},\,
k_\lambda\xr{\fm{\lambda}}\}$. The constraint equations (related to 
a polarization subalgebra ${\cal T}_p$ of 
$\tilde{{\cal T}}$) eliminate the 
$\chipm{}$ dependence of wave functions in (\ref{wavemass}) and keep 
$6=10-4$ field degrees of freedom for the symmetric case 
(massive spin 2 particle 
+ massive scalar field=trace of the symmetric tensor), 
and $3=6-3$ field degrees of freedom for 
the anti-symmetric case (massive pseudo-vector particle).

\section{Group approach to quantization of non-Abelian Yang-Mills theories}

Once we know how GAQ works on Abelian gauge field theories, let us 
tackle the non-Abelian case,  the underlying structure of which is similar 
to the previous case. However, new richness and subtle distinctions are 
introduced because of the non-Abelian character of the constraint subgroup 
$T={\rm Map}(\Re^4,{\bf T})=\{U(x)=e^{\varphi_a(x)T^a}\}$, where 
${\bf T}$ is some 
non-Abelian, compact Lie-group with Lie-algebra commutation relations  
$[T^a,T^b]=C^{ab}_c T^c$. 

As for the Abelian case, the traditional gauge transformation properties, 
\be
U(x)\rightarrow U'(x)U(x)\,,\;\;\;\;{\cal A}_\nu(x)\rightarrow 
U'(x){\cal A}_\nu(x)U'(x)^{-1}+U'(x)\partial_\nu U'(x)^{-1},\label{transf1}
\ee
for Lie-algebra valued vector potentials 
${\cal A}^\nu(x)= r_a^b {\cal A}^\nu_b(x) T^a$ 
($r_a^b$ denotes a coupling-constant matrix) have to be revised in order to
adapt it to an 
action of a group on itself. The solution for this is to
consider $A_\nu\equiv {\cal A}_\nu-U\partial_\nu U^{-1}$, which 
transforms homogeneously under the adjoint action of 
$T$, whereas the non-tensorial part $ U'(x)\partial_\nu U'(x)^{-1}$  
modifies the phase $\zeta$ of the wavefunctional 
$\Psi$ according to:
\bea
U(x)\rightarrow U'(x)U(x)\,,\;\;\;\;A_\nu(x)\rightarrow 
U'(x)A_\nu(x)U'(x)^{-1}\, ,\nn\\
\zeta\rightarrow \zeta \exp\left\{\frac{i}{r^2}
\int_\Sigma{ d\sigma_\mu(x) 
\,{\rm tr}\left[ U'(x)^{-1}\partial_\nu 
U'(x)\overpmup A^\nu(x)\right]}\right\}\,, 
\label{connection2}
\eea  
where we are restricting ourselves, 
for the sake of simplicity, 
to special unitary groups ${\bf T}$, so that the 
structure constants $C^{ab}_c$ are 
totally anti-symmetric, and the anti-hermitian generators $T^a$ can be chosen 
such that the Killing-Cartan metric is just $\tr (T^aT^b)=-\um\delta^{ab}$. 
For simple groups, the coupling-constant matrix 
$r_a^b$ reduces to a multiple of the identity $r_a^b=r\delta_a^b$, 
and we have $A^\mu_a=-\f{2}{r}\tr (T^a A^\mu)$. As above, 
the argument  
of the exponential in (\ref{connection2}) can be considered 
to be a piece of a two-cocycle $\xi:G\times G\rightarrow \Re$ 
constructed through a conserved current,  
$\xi(g'|g)=\int_\Sigma{ d\sigma_\mu(x) 
J^\mu(g'|g)(x)},\,\,\, g',g\in G$, so that it does not depend on the 
chosen spacelike hypersurface $\Sigma$. Let us also discard the 
Poincar\'e subgroup $P$ in the 
group $G$; that is, we shall consider, 
roughly speaking, $G=G_A\times_s T$, i.e. the semidirect product of 
the Abelian group $G_A$ of Lie-valued vector potentials and the  
group $T$. The reason 
for so doing is something more than a matter of symplicitly. 
In fact, we could make 
the kinematics ($P$) and the constraints ($T$) compatible 
at the price of introducing an infinite number of extra generators in the 
enveloping algebra of $G_A$, in a way that makes the 
quantization procedure quite unwieldy. We could also introduce a free-like 
(Abelian) kinematics without the need for extra generators at the price of 
appropriately adjusting the constraint set and the kinematics into a stable 
system, that is, by introducing secondary constraints. Nevertheless, unlike 
other standard approaches 
to quantum mechanics, GAQ still holds, even in the absence of a well-defined 
(space-)time evolution, an interesting and desirable 
property concerning the quantization of gravity (see, for example, 
\cite{Rovelli2}). The {\it true} dynamics [which preserves 
the constraints (\ref{tequiv})] will eventually arise as part of 
the set of {\it good} operators of the theory (see below).

We shall adopt a non-covariant approach and choose a $t={\rm constant}$ 
$\Sigma$-hypersurface to write the cocycle $\xi$. Also, we shall make 
partial use of the gauge freedom to set the temporal component $A^0=0$, so 
that the electric field is simply $\vec{E}_a=-\partial_0{\vec{A}}_a$ 
[from now on, 
and for the sake of simplicity,  we shall put any three-vector $\vec{A}$ as 
$A$, and we will understand $AE=\sum_{j=1}^3A^jE^j$, in the hope that no 
confusion will arise]. In this case, there is still a residual gauge 
invariance $T={\rm Map}(\Re^3,{\bf T})$ (see \cite{Jackiw}).

Taking all of this into account,  the explicit group law 
$\tg''=\tg'*\tg$ [with $\tg=(g;\zeta)=(A,E,U;\zeta)$] for the proposed 
infinite-dimensional {\it Yang-Mills quantizing group} $\TG$ is:

\bea
U''(x)&=&U'(x)U(x)\,,\nn\\
A''(x)&=&A'(x)+U'(x)A(x)U'(x)^{-1}\,,\nn\\
E''(x)&=&E'(x)+U'(x)E(x)U'(x)^{-1}\,,\nn\\
\zeta''&=&\zeta'\zeta\exp\left\{-\frac{i}{r^2}\sum_{j=1}^2
\xi_j(A',E',U'|A,E,U)\right\}\,;
\label{ley}\\
\xi_1(g'|g)&\equiv& \int{{d}^3x\,\tr\l[\,\left(\ba{cc} A' & 
E'\ea\right) S \left(\ba{c} U'AU'^{-1} \\ 
U'EU'^{-1} \ea\right)\r]}\nn\\
\xi_2(g'|g)&\equiv& \int{{d}^3x\,\tr\l[\,\left(\ba{cc} \nabla U'U'^{-1} & 
E'\ea\right) S \left(\ba{c} U'\nabla UU^{-1}U'^{-1} \\ 
U'EU'^{-1} \ea\right)\,\r]}\,,\nn 
\eea
\ni where $S=\left(\ba{cc} 0 & 1 \\ -1 & 0\ea\right)$ is a symplectic 
matrix. As above, the first cocycle  $\xi_1$ 
is meant to provide {\it dynamics} 
to the vector potential, so that the couple $(A,E)$ 
corresponds to a canonically-conjugate pair of coordinates,  
and $\xi_2$ is the non-covariant analogue of (\ref{connection2}). 
It is noteworthy that, unlike 
the Abelian case $T={\rm Map}(\Re^4,U(1))$,  
the simple (and non-Abelian) 
character of $T$ for the present case precludes a non-trivial central 
extension $\TT$ of $T$ by $U(1)$ given by the cocycle $\xi_3$ [see 
(\ref{3cocys})]. However, 
mass will enter the non-Abelian Yang-Mills theories through 
{\it pseudo-cocycles} [in fact, coboundaries 
$\xi_\lambda(g'|g)=\eta_\lambda(g'*g)-\eta_\lambda(g')-
\eta_\lambda(g)$, 
where $\eta_\lambda(g)$ is the generating function (of the coboundary)] 
with `mass' parameters $\lambda_a=m_a^3$, which 
define trivial extensions as such, but provide new commutation relations 
in the Lie algebra of $\TG$ and alter the number of degrees of freedom 
of the theory (see below). To make more explicit the intrinsic 
significance of the two quantities $\xi_j\,,\,\, j=1,2$, let us 
calculate the non-trivial Lie-algebra commutators of the right-invariant 
vector fields  from  the group law (\ref{ley}). 
They are explicitly:
\bea
\l[ \xr{A_a^j(x)}, \xr{E_b^k(y)}\r]&=&-\delta^{ab}\delta_{jk}
\delta(x-y)\Xi\,,\nn\\ 
\l[ \xr{E_a(x)}, \xr{\varphi_b(y)}\r]&=&-C^{ab}_c\delta(x-y)\xr{E_c(x)}+ 
\f{1}{r}\delta^{ab}\nabla_x\delta(x-y)\Xi\,,\label{commutators}\\ 
\l[ \xr{A_a(x)}, \xr{\varphi_b(y)}\r]&=&-C^{ab}_c\delta(x-y)\xr{A_c(x)}\nn\\ 
\l[ \xr{\varphi_a(x)}, \xr{\varphi_b(y)}\r]&=& 
-C^{ab}_c\delta(x-y)\xr{\varphi_c(x)}\,,\nn 
\eea 
which agree with those of Ref. \cite{Jackiw} when the identification  
$\hat{E}_a\equiv i\xr{A_a}\,,\,\,\hat{A}_a\equiv
i\xr{E_a}\,,\,\,\hat{G}_a\equiv i\xr{\varphi_a}$  is made [note that 
$\xr{A_a}\sim \f{\delta}{\delta A_a}$ and $\xr{E_a}\sim
 \f{\delta}{\delta E_a}$ near the identity element $\tg=e$ of $\TG$, which
motivates this particular identification]. 

Let us construct the Hilbert 
space of the theory. As already mentioned, the representation 
$L_{\tg'}\Psi(\tg)=\Psi(\tg'*\tg)$ of $\TG$ on $\TT$-equivariant wave 
functions (\ref{tequiv}) proves to be reducible. The reduction is 
achieved by means of polarization 
conditions (\ref{polcond})   which 
contain finite right-transformations generated by left-invariant 
vector fields $\xl{}$ devoid of dynamical content (that is, without 
a canonically conjugated counterpart), and half of the left-invariant vector 
fields related to dynamical coordinates 
(either ``positions'' or ``momenta''). 
The left-invariant vector fields without a canonically conjugated counterpart 
are the combinations 
\be
{\cal G}_c=<\xl{\theta_a}
\equiv\xl{\varphi_a}-\f{1}{r}\nabla\cdot\xl{A_a}>\,,\label{cara}
\ee 
which close a Lie subalgebra isomorphic to the {\it higher-order} 
gauge subalgebra (it generates a horizontal ideal of the right-enveloping 
algebra ${\cal U}(\tilde{{\cal G}}^R)$ and proves to be zero on polarized 
wave functionals)
\be
{\cal G}_{{\small{\rm gauge}}}= <\xr{\phi_a(x)}=\xr{\varphi_a(x)}+
\frac{1}{r}\nabla\cdot\xr{A_a(x)}
+iC^{ab}_c\xr{A_b(x)}\cdot\xr{E_c(x)}>\,.\label{idealHO}
\ee
As for the Abelian case, the flow of (\ref{idealHO}) recovers 
the time-independent (residual) transformation (\ref{transf1}).

The characteristic subalgebra ${\cal G}_c$ can be completed to a 
 full polarization subalgebra ${\cal G}_p$ in two different ways:
\be
{\cal G}_p^{(A)}\equiv <{\cal G}_c,\,\,\xl{A_b}\,\,
\forall b>,\;\;\;
{\cal G}_p^{(E)}\equiv <{\cal G}_c,\,\,\xl{E_b}\,\,
\forall b>,
\ee
each one giving rise to a different representation space: a) the electric 
field representation and b) the magnetic field representation, respectively.
The polarized, $U(1)$-equivariant functions are:
\bea
\xl{}\Psi_A=0,\,\,\forall \xl{}\in {\cal G}_p^{(A)}&\Rightarrow &
\Psi_A(A,E,U;\zeta)=\zeta e^{-\coci{AE-U\nabla U^{-1}E}}
\Phi_A(E)\,,\nn\\
\xl{}\Psi_E=0,\,\,\forall \xl{}\in {\cal G}_p^{(E)}&\Rightarrow &
\Psi_E(A,E,U;\zeta)=\zeta e^{\coci{AE-U\nabla U^{-1}E}}
\Phi_E({\cal A})\,,\label{polwave}
\eea
where $\Phi_A$ and $\Phi_E$ are arbitrary functionals of $E$ and 
${\cal A}\equiv A+\nabla UU^{-1}$, respectively. The trivial fibration of 
$\TT=T\times U(1)$ for the massless case allows us to impose the whole 
$T_p=T$ group as constraint conditions $L_{\tit'}\Psi(\tg)=\Psi,\,\forall 
\tit'=(0,0,U';1)\in T$ on wavefunctionals. For each representation space, 
the constraint conditions read: 
\bea
L_{\tit'}\Psi_A(\tg)=\Psi_A(\tg)&\Rightarrow &
\Phi_A(E)=e^{-2\coci{{U'}^{-1}\nabla U'E}}
\Phi_A(U'E{U'}^{-1})\,,\nn\\
L_{\tit'}\Psi_E(\tg)=\Psi_E(\tg)&\Rightarrow &
\Phi_E({\cal A})=\Phi_E(U'{\cal A}{U'}^{-1}+\nabla U'{U'}^{-1})\,,\label{liga}
\eea
which are the finite counterpart of the infinitesimal, 
quantum implementation  
of the Gauss law $\xr{\varphi_a(x)}\Psi=-i\hat{G}_a(x)\Psi=0$. 
The polarized and constrained  wavefunctionals (\ref{polwave},\ref{liga}) 
define the constrained Hilbert space ${\cal H}_{{\small{\rm phys}}}(\TG)$ 
of the theory, and the 
infinitesimal form $\xr{\tg}\Psi(\tg)$ of the finite left-action 
$L_{\tg'}\Psi(\tg)$ of $\TG$ on ${\cal H}(\TG)$ provides the action of the 
operators $\hat{A}_a,\hat{E}_a,\hat{G}_a$ on wave functions 
(see \cite{ym2} for more details). 

 The good operators  
$\tilde{{\cal G}}_{{\tiny {\rm good}}}$ for this case 
have to be found inside 
the right-enveloping algebra 
${\cal U}(\tilde{{\cal G}}^R)$ 
of polynomials of the basic operators $\hat{A}_a(x),\,\hat{E}_b(x)$, as 
forming part of the {\it normalizer} of $T$ (see Eq. (\ref{gooddef})). 
In particular, some good operators are:
\be
\tilde{{\cal G}}_{{\rm {\small good}}}=<\tr [ 
\hat{E}^j(x)\hat{B}^k(x)],\,\tr [ 
\hat{E}^j(x)\hat{E}^k(x)],\, \tr [ \hat{B}^j(x)\hat{B}^k(x)],\,
\Xi>\,,\label{good}
\ee
\ni  where $\hat{B}_a\equiv \nabla\wedge\hat{A}_a-\um r C^{ab}_c
 \hat{A}_b\wedge\hat{A}_c$ (the magnetic field)
can be interpreted as a ``correction'' to $\hat{A}_a$   
that, unlike $\hat{A}_a$, transforms homogeneously under the adjoint
action of $T$ [see 2nd line of (\ref{commutators})]. The components 
$\hat{\Theta}^{\mu\nu}(x)$ of the 
standard canonical energy-momentum tensor for Yang-Mills theories  
are linear combinations of operators in (\ref{good}); for example,  
$\hat{\Theta}^{00}(x)=-\tr [ E^2(x)+B^2(x)]$ is the Hamiltonian 
density. In this way, Poincar\'e invariance is retrieved in the 
constrained theory. 

Let us mention, for the sake of completeness, that the actual use
of good operators is not restricted to first- and second-order operators. 
Higher-order operators can constitute a useful tool in finding the whole 
constrained Hilbert space ${\cal H}_{{\small{\rm phys}}}(\TG)$. In fact, it 
can be obtained from a constrained (physical) state $\Phi^{(0)}$, 
i.e. $\hat{G}_a\Phi^{(0)}=0$,
on which the energy-momentum tensor has null expectation value 
$\langle \Phi^{(0)}|\hat{\Theta}^{\mu\nu}|\Phi^{(0)}\rangle=0$, by taking the 
orbit of the rest of good operators passing through this ``vacuum''. This has 
been indeed a rather standard technique (the Verma module approach) in 
theories where null vector states are present in the original Hilbert space 
\cite{Kac,Feigin,virazorro}. From an other point of view, with regard to  confinement,
exponentials of the form $\varepsilon_{\Sigma_2}
\equiv\tr\left[\exp(\epsilon_{jkl}\int_{\Sigma_2}{d\sigma^{jk}\hat{E}^l})
\right]$ and $\beta_{\Sigma_2}\equiv\tr\left[\exp(\epsilon_{jkl}
\int_{\Sigma_2}{d\sigma^{jk}\hat{B}^l})\right]$, 
where ${\Sigma_2}$ is a two-dimensional surface in three-dimensional space, 
are good operators related to Wilson loops. 

As a previous step before examining the massive case, 
let us show how new physics can enter the theory by considering 
non-trivial representations $D_{\TT}^{(\epsilon)}$ of $\TT$ or, in an 
equivalent way, by introducing certain extra coboundaries in the 
group law (\ref{ley}). Indeed, more general representations 
for the  constraint 
subgroup $T$, namely the one-dimensional representation 
$D_{\TT}^{(\epsilon)}(U)=e^{i\epsilon_{{}_U}}$,  
can be considered if we impose 
additional boundary conditions like $U(x)
\stackrel{x\rightarrow\infty}{\longrightarrow}\pm I$; this means that we 
compactify the space $\Re^3\rightarrow S^3$, so that the group $T$ 
fall into disjoint homotopy classes $\{U_l\,,\,\epsilon_{{}_{U_l}}=
l\vartheta\}$ labelled by integers $l\in Z=\Pi_3({\bf T})$ 
(the third homotopy group). 
The index $\vartheta$ (the {\it $\vartheta$-angle} 
\cite{Jackiwtheta}) parametrizes 
{\it non-equivalent quantizations}, as the Bloch momentum $\epsilon$ does  
for particles in periodic potentials, where the wave function acquires 
a phase $\psi(q+2\pi)=e^{i\epsilon}\psi(q)$ after a translation of, 
let us say, $2\pi$. 
The phenomenon of non-equivalent quantizations can also be reproduced by 
keeping the  constraint condition  $D_{\TT}^{(\epsilon)}(U)=1$, as in 
(\ref{liga}),  
at the expense of introducing a new (pseudo) cocycle 
$\xi_\vartheta$ which is added to the 
previous cocycle $\xi=\xi_1+\xi_2$ in (\ref{ley}). The generating function 
of $\xi_\vartheta$ is $\eta_\vartheta(g)=\vartheta\int{d^3x\, 
{\cal C}^0(x)}$, 
where ${\cal C}^0$ is 
the temporal component of the {\it Chern-Simons secondary characteristic 
class}
\be
{\cal C}^\mu=-\frac{1}{16\pi^2}\epsilon^{\mu\alpha\beta\gamma}{\rm tr}
({\cal F}_{\alpha\beta}{\cal A}_\gamma-\frac{2}{3}{\cal A}_\alpha 
{\cal A}_\beta {\cal A}_\gamma)\,,
\ee
which is the vector whose divergence equals the Pontryagin density  
 ${\cal P} = \partial_\mu {\cal C}^\mu = -\frac{1}{16\pi^2} 
{\rm tr} \break ({}^*{\cal F}^{\mu\nu} {\cal F}_{\mu\nu})$ 
(see \cite{Jackiw}, for instance). 
Like some total derivatives (namely, the Pontryagin density), 
which do not modify 
the classical equations of motion when added to the Lagrangian but have a 
non-trivial effect in the quantum theory, the coboundaries 
$\xi_\vartheta$ give rise to non-equivalent quantizations parametrized 
by $\vartheta$ when 
the topology of the space is affected by the imposition of 
certain boundary conditions (``compactification of the space''), 
even though they are trivial cocycles of  the ``unconstrained'' theory. 
The phenomenon of non-equivalent quantizations can also be
understood  sometimes  as a {\it Aharonov-Bohm-like effect} (an effect 
experienced by the quantum particle but not by the classical particle) 
and $\delta\eta(g)=
\frac{\delta\eta(g)}{\delta g^j}\delta g^j$ can be seen 
as an {\it induced gauge connection} (see \cite{FracHall} for the example 
of a superconducting ring threaded by a magnetic flux) which modify the 
momenta according to minimal coupling. 

There exist other kind of coboundaries  
generated by functions $\eta(g)$ with non-trivial gradient 
$\left.\delta\eta(g)\right|_{g=e}\not=0$ at the identity $g=e$,  
which provide a contribution to the connection form of the theory 
(\ref{thetagen}) and the 
structure constants of the original Lie algebra (\ref{commutators}). We 
shall call these {\it pseudo-cocycles}, since they give rise to 
{\it pseudo-cohomology classes} related with coadjoint orbits of semisimple 
groups \cite{Marmo}. Whereas coboundaries generated by  
global functions on the original 
(infinite-dimensional) group $G$ having trivial gradient at the identity, 
namely $\xi_\vartheta$, contribute the quantization 
with global (topological) effects, pseudo-cocycles can give dynamics 
to some non-dynamical operators and provide 
new couples of conjugated field operators, thus 
substantially modifying the theory. Let us see how, in fact, 
the possibility of {\it non-Abelian} representations of $\TT$ is equivalent 
to the introduction of 
new pseudo-cocycles in the centrally extended group law (\ref{ley}).

\subsection{Cohomological origin of mass and alternatives to the Higgs 
mechanism}

Non-trivial transformations of the wavefunctional 
$\Psi$ under the action of $T$ can also be reproduced by considering 
the pseudo-cocycle 
\be
\xi_\lambda(g'|g)\equiv -2\int{{d}^3x\, {\rm tr}[ 
\lambda\left(\log (U'U)-\log U'-\log U\right)]}\,,\nn
\ee
which is generated by $\eta_\lambda(g)=-2\int{{d}^3x\, {\rm tr}[
\lambda\log U]}$, where $\lambda=\lambda_aT^a$ is a matrix carrying some 
parameters $\lambda_a$ (with mass-cubed dimension) 
which actually characterize the representation of 
$\TG$. However, unlike $\xi_\vartheta$, this pseudo-cocycle (whose 
generating function $\eta_\lambda$ has a non-trivial gradient at the identity 
$g=e$) alters the Lie-algebra commutators of $T$ 
and leads to the appearance of new central terms at the right-hand side 
of the last line of Eq. (\ref{commutators}). More explicitly:
\be
\l[ \xr{\varphi_a(x)}, \xr{\varphi_b(y)}\r]= 
-C^{ab}_c\delta(x-y)\xr{\varphi_c(x)}
-C^{ab}_c\frac{\lambda^c}{r^2}\delta(x-y)\Xi\,. \label{masin}
\ee
The appearance of new 
central terms proportional to the  parameter $\lambda^c$ at the 
right-hand side of (\ref{masin})  restricts the number 
of vector fields in the characteristic subalgebra (\ref{cara}), which 
now consists of 
\be
{\cal G}_c=<
\xl{\theta_a}\,\,/\,\,C^{ab}_c\lambda^c\not=0\,\,
\forall b>
\ee
(that is, the subalgebra of non-dynamical generators), with 
respect to the case $\lambda^c=0$, where ${\cal G}_c$ is isomorphic to 
${\cal T}$. Therefore, the pseudo-cocycle $\xi_\lambda$ 
provide new degrees of freedom to the theory; that is, new 
pairs of generators $(\xr{\varphi_a},\xr{\varphi_b})$,  
with $C^{ab}_c\lambda^c\not=0$, become 
conjugated and, therefore, new basic field operators 
enter the theory. Let us see how these new degrees of freedom 
are transfered to the 
vector potentials to conform {\it massive} vector bosons with mass cubed 
$m_a^3=\lambda_a$.

In order to count the number 
of degrees of freedom for a given structure subgroup $\TT$ and a given 
``mass'' matrix $\lambda$, let us denote by 
$\tau={\rm dim}({\bf T})$ and $c={\rm dim}({\bf G}_c)$ the dimensions of 
the rigid subgroup of $T$ and 
the characteristic subgroup $G_c$, respectively. 
In general, for an arbitrary mass matrix 
$\lambda$, we have $c\leq\tau$. Unpolarized, $U(1)$-equivariant functions 
$\Psi(A^j_a,E^j_a,\varphi_a)$ depend on $n=2\times 3\tau+\tau$ field 
coordinates in $d=3$ spatial dimensions; 
polarization equations introduce $p=c+
\frac{n-c}{2}$ independent restrictions on 
wave funtions, corresponding to $c$ non-dynamical coordinates in $G_c$ and 
half of the dynamical ones; finally, constraints provide 
$q=c+\frac{\tau-c}{2}$ additional restrictions which leave 
$f=n-p-q=3\tau-c$ field degrees of freedom (in $d=3$). 
Indeed, for the massive case, constraints are 
{\it second-class} and we can only impose a polarization subalgebra 
${\cal T}_p\subset\tilde{{\cal T}}$, which contains a 
characteristic subalgebra 
${\cal T}_c=<\xr{\varphi_a}\,\,/\,\, 
C^{ab}_c\lambda^c=0\,\,\forall b>
\subset \tilde{{\cal T}}$ (which is isomorphic to ${\cal G}_c$) 
and half of the rest of generators 
in $\tilde{{\cal T}}$ (excluding $\Xi$); that is, 
$q=c+\frac{\tau-c}{2}\leq \tau$ independent 
constraints which lead to constrained wave functions having support on 
$f_{m\not=0}=2c+3(\tau-c)\geq f_{m=0}$ arbitrary fields corresponding to 
$c$ massless vector bosons attached to ${\cal T}_c$ and 
$\tau-c$ massive vector bosons.  In particular, for the massless 
case we have ${\cal T}_c={\cal T}$, i.e. $c=\tau$,  
since constraints are {\it first-class} 
(that is, we can impose $q=\tau$ restrictions) and constrained wave 
functions have support on $f_{m=0}=3\tau-\tau=2\tau\leq f_{m\not=0}$ 
arbitrary fields corresponding to $\tau$ massless vector bosons. 
The subalgebra ${\cal T}_c$ corresponds to the unbroken 
gauge symmetry of the constrained theory and proves to be an  
{\it ideal} of $\tilde{{\cal G}}_{\small{\rm good}}$ [remember 
the characterization of {\it good} operators before Eq. (\ref{good})].

Let us work out a couple of examples. Cartan (maximal Abelian) 
subalgebras of $T$ will be privileged as candidates for the unbroken 
electromagnetic gauge symmetry. Thus, let us use the Cartan basis  
$<H_i,E_{\pm \alpha}>$ instead of $<T^a>$, and denote 
$\{\varphi_i,\varphi_{\pm \alpha}\}$ the coordinates of $T$ 
attached to this basis (i.e. $\varphi_{\pm \alpha}$ are complex field 
coordinates attached to each root $\pm\alpha$ and $\varphi_i$ are 
real field coordinates attached to the maximal torus of {\bf T}). 
For $T={\rm Map}(\Re^3,SU(2))$ and 
$\lambda=\lambda_1H_1$, the 
characteristic, polarization and constraint subalgebras (leading 
to the electric field representation) are:

\be
{\cal G}_c=<\,\xl{\theta_1}\,>,\,\,\,
{\cal G}_p^{(A)}=<\,\xl{\theta_1},\xl{\theta_{+1}},\xl{A}\,>,\,\,\,
{\cal T}_p=<\,\xr{\varphi_1},\xr{\varphi_{-1}}\,>\,.
\ee
which corresponds to a self-interacting theory of a massless vector 
boson $A_1$ [with unbroken gauge subgroup 
$T_c={\rm Map}(\Re^3,U(1))\subset {\rm Map}(\Re^3,SU(2))$] and 
two charged vector bosons $A_{\pm 1}$ with mass cubed $m_1^3=\lambda_1$. For 
$T={\rm Map}(\Re^3,SU(3))$ and $\lambda=\lambda_2 H_2$, we have
\be\ba{cc}
{\cal G}_c=<\,\xl{\theta_{1,2}},\xl{\theta_{\pm 1}}\,>,\,\,\,
{\cal G}_p^{(A)}=<\,\xl{\theta_{1,2}},\xl{\theta_{\pm 1}},
\xl{\theta_{+2,+3}},\xl{A}\,>,\\ 
{\cal T}_p=<\,\xr{\varphi_{1,2}},\xr{\varphi_{\pm 1}},
\xr{\varphi_{-2,-3}}\,>\ea\,.
\ee
Thus, the constrained theory corresponds to a 
self-interacting theory of two massless vector 
bosons $A_{1,2}$, two massless charged vector bosons $A_{\pm 1}$ 
[the unbroken gauge subgroup is now $T_c={\rm Map}(\Re^3,SU(2)\times U(1))$] 
and four charged vector bosons $A_{\pm 2,\pm 3}$ with mass cubed 
$m^3_2=\lambda_2$. For $SU(N)$ we have several symmetry breaking patterns 
related to the different choices of mass matrix 
$\lambda=\sum_{i=1}^{N-1}\lambda_i H_i$.

Summarizing, new basic operators 
$\hat{G}_{\pm \alpha}\equiv \xr{\varphi_{\pm\alpha}}$,  
with $C^{\alpha-\alpha}_i\lambda^i\not=0$,  and new {\it non-trivial}  
good operators $\hat{C}_i=\{$Casimir operators of $\TT\}$ 
($i$ runs the range of {\bf T}) enter the theory, in
contrast to the massless case. For example, for $T={\rm Map}(\Re^3,SU(2))$, 
the Casimir operator is 
\be
\hat{C}(x)= (\hat{G}_1(x)+\frac{\lambda_1}{r^2})^2+
2(\hat{G}_{+1}(x)\hat{G}_{-1}(x)
+\hat{G}_{-1}(x)\hat{G}_{+1}(x))\,.
\ee 
Also, the Hamiltonian density 
$\hat{\Theta}^{00}(x)=-\tr\left[ E^2(x)+B^2(x)\right]$ for $m=0$ is  
affected in the massive case $m\not=0$ by the presence of extra terms 
proportional to these non-trivial Casimir operators (which are zero on 
constrained wave functionals in the massless case), as follows:
\be
\hat{\Theta}^{00}_{m\not=0}(x)=\hat{\Theta}_{m=0}^{00}(x)+\sum_i{
\frac{r^2}{m_i^2}\hat{C}_i(x)}\,.
\ee
Thus, the Sch\"odinger equation 
$\int{d^3x\hat{\Theta}^{00}_{m\not=0}(x)}\Phi={\cal E}\Phi$ is also 
modified by the presence of extra terms.

\section{Conclusions and outlooks}

We have seen how the appearance of (quantum) central terms in the Lie-algebra 
of symmetry of gauge theories provides new degrees of freedom which are 
transferred to the potentials to conform massive bosons. Thus, 
the appearance of mass seems to have a {\it cohomological origin}, beyond any 
introduction of extra (Higgs) particles. Nevertheless, the introduction of 
mass through the pseudo-cocycle $\xi_\lambda$ is equivalent to the choice of a 
vacuum in which some generators of the unbroken gauge symmetry $T_c$ have a 
non-zero expectation value proportional to the mass parameter 
(see \cite{ym2}). This fact reminds us of the Higgs mechanism 
in non-Abelian gauge theories, where the Higgs fields point to the direction 
of the non-null vacuum expectation values. However, the spirit of this 
standard approach to supply mass and the one explained in this paper are 
radically different, even though they have some characteristics in common.
In fact, we are not making use of extra scalar fields in the theory to 
provide mass to the vector bosons, but it is the gauge group itself 
which acquires dynamics for the massive case and transfers degrees of freedom 
to the vector potentials.

It is also worth devoting some words with regard to renormalizability for
the case of a non-trivial mass matrix $\lambda\neq 0$. Obviously we should 
refer to finitness simply, since we are dealing with a non-perturbative
formulation. But the major virtue of a group-theoretic algorithm is that we 
automatically arrive at normal-ordered, finite quantities, and this is true 
irrespective of the ``breaking'' of the symmetry. We must notice that when we 
use the term `unbroken gauge symmetry', in referring to $T_c$, we
mean only that this subgroup of $\TT$ is devoid of dynamical content; the gauge group 
of the constrained theory is, in both the massless and massive cases, 
the group $T=\TT/U(1)$ although, for the massive case, only a polarization 
subgroup $T_p$ can be consistently imposed as a constraint. This is also 
the case of the Virasoro algebra 
\be
[\hat{L}_n,\hat{L}_m]=(n-m)\hat{L}_{n+m}+
\frac{1}{12}(cn^3-c'n)\delta_{n,-m}\hat{1}\,,\label{viral}
\ee 
in String Theory, where 
the appearance of central terms does not spoil gauge invariance but 
forces us to impose half of the 
Virasoro operators only (the positive modes $\hat{L}_{n\geq 0}$) 
as constraints. In general, although diffeomorphisms usually appear as 
a constriant algebra under which one might expect all the physical states 
to be singlets, dealing with them in the 
quantum arena, the possibility of central extensions should be 
naturally allowed and welcomed, as they provide new richness to the theory.

Pseudo-cocycles similar to $\xi_\lambda$ do also appear in the
representation of Kac-Moody and conformally invariant theories in general, 
although the pseudo-cocycle parameters are usually hidden in a redefinition of 
the generators involved in the pseudo-extension (the argument of the 
Lie-algebra generating function). For the Virasoro algebra, the redefinition 
of the $\hat{L}_0$ generator produces a non-trivial expectation 
value in the vacuum $h\equiv (c-c')/24$ \cite{virazorro}.

Fermionic matter can also be 
incorporated into the theory through extra (Dirac fields) coordinates 
$\psi_l(x),\, l=1,\dots, p$ 
and an extra two-cocycle  
\be
\xi_{{\small{\rm matter}}}= i\int{{d}^3x
\,\left(\bar{\psi}' \gamma^0\rho(U')\psi-\bar{\psi}\rho(U'^{-1}) 
\gamma^0\psi'\right)}\,,
\ee
where $\rho(U)$ is a $p$-dimensional representation of {\bf T} acting on 
the column vectors $\psi$, and $\gamma^0$ is the time component of 
the standard Dirac matrices $\gamma^\mu$ (see \cite{ym2} for more details).

\section*{Acknowledegment}

M. Calixto thanks the University of Granada for a Postdoctoral grant and the 
Department of Physics of Swansea for its hospitality.

\end{document}